\documentclass[
reprint,
superscriptaddress,
amsmath,amssymb,
aps,
prb,
]{revtex4-2}

\usepackage{graphicx}
\usepackage{dcolumn}
\usepackage{bm}
\usepackage{subcaption}
\usepackage{qcircuit}
\usepackage{braket}

\begin{document}
\title{Tuning the Quantum Mpemba Effect in Isolated System by Initial State Engineering}

\author{Yihan Yu}
\affiliation{Beijing National Laboratory for Condensed Matter Physics, Institute of Physics, Chinese Academy of Sciences, Beijing 100190, China}
\affiliation{School of Physical Sciences, University of Chinese Academy of Sciences, Beijing 100049, China}

\author{Tianren Jin}
\affiliation{Beijing National Laboratory for Condensed Matter Physics, Institute of Physics, Chinese Academy of Sciences, Beijing 100190, China}
\affiliation{School of Physical Sciences, University of Chinese Academy of Sciences, Beijing 100049, China}

\author{Lv Zhang}
\affiliation{Beijing National Laboratory for Condensed Matter Physics, Institute of Physics, Chinese Academy of Sciences, Beijing 100190, China}
\affiliation{School of Physical Sciences, University of Chinese Academy of Sciences, Beijing 100049, China}

\author{Kai Xu}
\affiliation{Beijing National Laboratory for Condensed Matter Physics, Institute of Physics, Chinese Academy of Sciences, Beijing 100190, China}
\affiliation{School of Physical Sciences, University of Chinese Academy of Sciences, Beijing 100049, China}
\affiliation{Beijing Key Laboratory of Fault-Tolerant Quantum Computing, Beijing Academy of Quantum Information Sciences, Beijing 100193, China}
\affiliation{Hefei National Laboratory, Hefei 230088, China}
\affiliation{Songshan Lake Materials Laboratory, Dongguan, Guangdong 523808, China}

\author{Heng Fan}
\email{hfan@iphy.ac.cn}
\affiliation{Beijing National Laboratory for Condensed Matter Physics, Institute of Physics, Chinese Academy of Sciences, Beijing 100190, China}
\affiliation{School of Physical Sciences, University of Chinese Academy of Sciences, Beijing 100049, China}
\affiliation{Beijing Key Laboratory of Fault-Tolerant Quantum Computing, Beijing Academy of Quantum Information Sciences, Beijing 100193, China}
\affiliation{Hefei National Laboratory, Hefei 230088, China}
\affiliation{Songshan Lake Materials Laboratory, Dongguan, Guangdong 523808, China}
\date{2025-05-02}

\begin{abstract}
We investigate the quantum Mpemba effect (QME) in isolated, non-integrable quantum systems, where relaxation dynamics depend on structure of the initial states. By analyzing the distribution of initial states across symmetrical subspaces, we identify a tunable mechanism that influences the emergence of QME, showing faster relaxation from certain out-of-equilibrium states. Additionally, we propose an experimentally realizable quantum circuit, which requires no complex controls on quantum simulator platforms and serves to verify our theoretical predictions. These results establish symmetry-resolved state engineering as a practical tool for manipulating non-equilibrium quantum dynamics.
\end{abstract}
\maketitle

\section{Introduction}
The study of non-equilibrium quantum systems has garnered significant attention in recent years~\cite{teza2025speedups}. The QME, a non-equilibrium quantum analog of the classical Mpemba effect, describes a counterintuitive phenomenon where a system initially further from equilibrium relaxes faster to equilibrium than that of a system initially closer to equilibrium. This effect, first observed and verify in various classical systems~\cite{mpemba1969cool,PhysRevLett.119.148001,lu2017nonequilibrium,klich2019mpemba,kumar2020exponentially,bechhoefer2021fresh,teza2023relaxation,kumar2022anomalous,malhotra2024double}, e.g., the freezing of hot water faster than warm water, has now been extended to quantum regimes and investigated in various settings~\cite{aharony2024inverse,zhang2025observation,joshi2024observing,nava2024mpemba,chatterjee2023quantum,carollo2021exponentially,manikandan2021equidistant,kochsiek2022accelerating}. 

Previous studies have explored the QME from various point of views. For instance, in integrable systems~\cite{10.21468/SciPostPhys.15.3.089,PhysRevLett.133.010401,PhysRevX.12.011006}, the relaxation velocity has been linked to the slowest diffusive modes of the system. In Markovian frameworks~\cite{nava2019lindblad,graf2024role,PhysRevLett.133.140404}, the relaxation dynamics are connected to the different modes of the master equation governing evolution of the system. 

In addition to the studies conducted within integrable and Markovian frameworks, a diverse range of approaches have also been exploited to investigate the quantum Mpemba effect across various physical systems. For instance, non-Markovian frameworks have been considered to account for systems with finite memory times due to interactions between finite baths and the system~\cite{strachan2024non}. Furthermore, the effect has been analyzed in driven dissipative systems with oscillatory electric fields, revealing transitions between real and complex eigenvalues that influence anomalous relaxation~\cite{chatterjee2024multiple,kheirandish2024mpemba}. Perturbative methods~\cite{ivander2023hyperacceleration} and alternative dynamical approaches, such as those based on Davis maps~\cite{moroder2024thermodynamics}, have also been proposed. Information geometry has been employed to quantify the effect~\cite{bettmann2025information}, while connections to quantum chaos have been explored in the context of the SYK model~\cite{wang2024mpemba}. Moreover, the quantum Mpemba effect has also been investigated in many-body localized systems~\cite{liu2024quantum}. 

While significant progress has been made in understanding the QME in specific models and regimes, the exploration of general quantum systems particularly those that are non-integrable or non-Markovian still leaves ample room for further investigation. Though work by Rylands \textit{et al.}~\cite{PhysRevLett.133.010401} reports that their theory could be extended to the non-integrable case, the concept of quasiparticle still needs to be induced. An alternative perspective on the origin of the QME that does not rely on integrability has been provided by Liu \textit{et al.}~\cite{PhysRevLett.133.140405}, whose insightful work employs a random circuit framework to demonstrate the emergence of the effect.

In this work, we address this open direction by developing a general framework for isolated quantum systems evolving under Schrödinger dynamics. We derive an analytical expression that captures the relaxation dynamics of such systems and validate our findings through numerical simulations of a non-integrable model. Furthermore, we reveal that the relaxation velocity is fundamentally governed by the spectral distribution of the system, which modulates entropy production and the efficiency of ergodic exploration in the effective dynamical landscape. 

This framework offers new insights into the intrinsic connections between quantum chaos, information scrambling, and thermodynamic behavior. Our results not only shedding a new light on understanding of the QME but also open practical pathways for engineering relaxation dynamics in quantum technologies, with potential applications in quantum information processing and quantum thermodynamic control.

\section{Setup}
To frame the discussion within the context of Schrödinger evolution, we begin by considering an isolated quantum system governed by the Hamiltonian \( \hat{H} \). In this setting, we focus on the relaxation dynamics of a non-equilibrium system, assuming that the system possesses a locally conserved charge \( \hat{Q} \) that commutes with the Hamiltonian, i.e., 
\begin{eqnarray}
[\hat{H}, \hat{Q}] = 0.
\end{eqnarray}
Let us further partition the system into two subsystems, \( A \) and \( \bar{A} \), such that the total charge \( \hat{Q} \) can be expressed as the sum of the charges in each subsystem:
\begin{eqnarray}
\hat{Q} = \hat{Q}_A + \hat{Q}_{\bar{A}}.
\end{eqnarray}
where \( \hat{Q}_A \) and \( \hat{Q}_{\bar{A}} \) are the charge operators acting on the subsystems \( A \) and \( \bar{A} \), respectively. We now turn to the typical scenario encountered in thermalization studies, where one subsystem is much smaller than the other. In this case, the smaller subsystem is referred to as the Quantum Open System (QOS), while the larger subsystem is termed the Thermal Bath (QTB).

\begin{figure}[]
\centering
\includegraphics[width=0.48\textwidth]{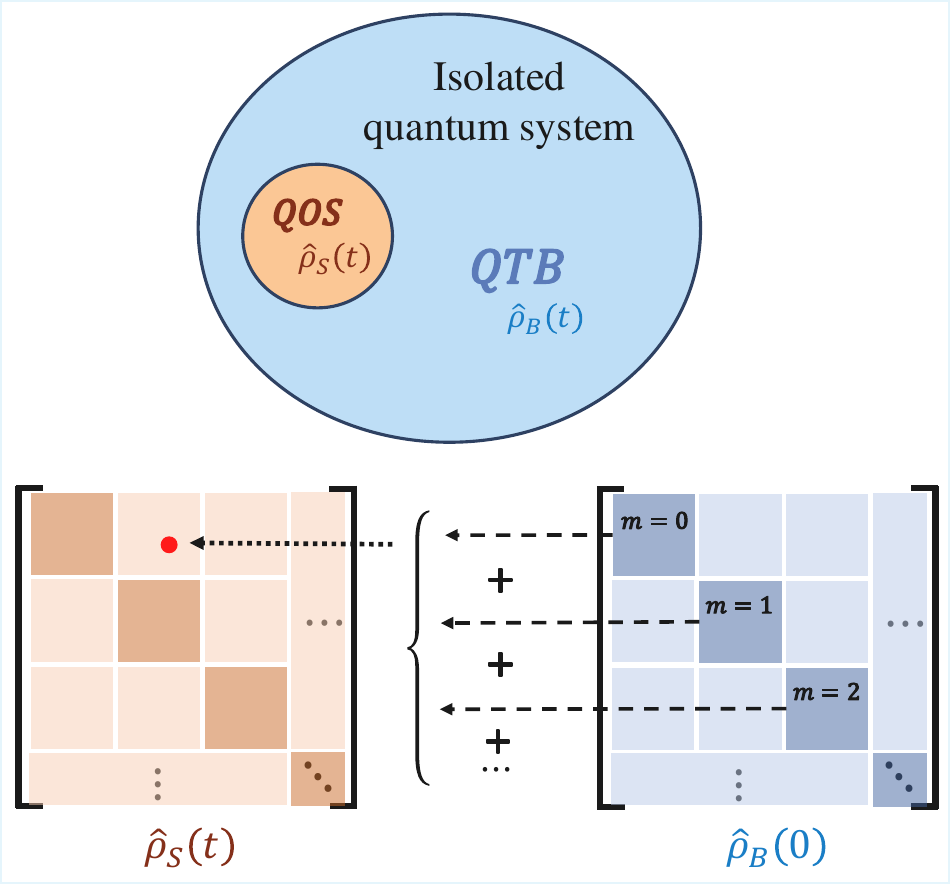}
\caption{Illustration of the theoretical setup.}
\label{fig:demo}
\end{figure}

Due to the presence of a locally conserved charge, our quantum system can be decomposed into distinct subspaces within its Hilbert space, whether considering the subsystem alone or the entire system. Specifically, under the partitioning into the Quantum Open System (QOS) and the Thermal Bath (QTB), as shown in Fig.~\ref{fig:demo} we can define the local charge operator for each subsystem as $\hat{Q} = \hat{Q}_S + \hat{Q}_B$, where $\hat{Q}_S$ and $\hat{Q}_B$ correspond to the charge operators acting on the QOS and QTB, respectively. Each state within a given subspace of the QOS/QTB can then be labeled as $\ket{S^q_i}$ or $\ket{B^q_i}$, where $q$ denotes the charge associated with that state and the label $i$ denotes a particular state within the $q$-th subspace. Formally, this means that $\hat{Q}_S \ket{S^q_i} = q \ket{S^q_i}$ and $\hat{Q}_B \ket{B^q_i} = q \ket{B^q_i}$.

Simultaneously, we can define the eigenstates of the total charge operator $\hat{Q}$ and the Hamiltonian $\hat{H}$ for the full system, denoted by $\ket{m,E_n}$, where the eigenvalue $m$ corresponds to the charge and $E_n^m$ is the corresponding eigenenergy. These states satisfy the relations
\begin{eqnarray}
\hat{Q} \ket{m,E_n} = m \ket{m,E_n}, \quad \hat{H} \ket{m,E_n} = E_n^m \ket{m,E_n}.
\end{eqnarray}

We denote the reduced density matrix of QOS as $\hat{\rho}_S = \text{Tr}_{B}\left(\ket{\Psi}\bra{\Psi}\right)$. Consider two basis states of the QOS, \( \ket{e_1} \) and \( \ket{e_2} \). The corresponding off-diagonal element is 
\begin{eqnarray}
\bra{e_1} \hat{\rho}_S \ket{e_2} = \bra{\Psi} \left(\ket{e_2} \bra{e_1} \otimes \hat{I}_B\right) \ket{\Psi} ,
\end{eqnarray}
where \( \hat{I}_B \) is the identity operator acting on the QTB. This expression captures the correlation between the states \( \ket{e_1}  \) and \( \ket{e_2} \) in the QOS, mediated by their interactions with the QTB. It is expected that in an thermalization process, QOS would relaxes to a stationary state~\cite{RevModPhys.83.863,Barthel2007Dephasing,Short2011Quantum,Cramer2007Exact,Reimann2012Equilibration,PhysRevLett.133.010401,fagotti2014relaxation,calabrese2016introduction,Vidmar_2016,Essler_2016,10.21468/SciPostPhysLectNotes.18,Bastianello_2022,Alba_2021,Fagotti_2014}, except some exotic instances~\cite{fagotti2016charges,bertini2015pre,10.21468/SciPostPhys.15.3.089,PhysRevLett.133.140405}. Considering the locally conserved charge our system has, we could find this stationary state restoring the sysmmetry corresponding to this charge, i.e., $\lim_{t\rightarrow\inf}[\hat{\rho}_S,\hat{Q}_S]=0$. Or under another view, the off-diagonal elements \( \bra{e_1} \hat{\rho}_A \ket{e_2} \) would generally decay to zero for $\ket{e_1},\ket{e_2}$ in different sysmmetry spaces of $\hat{Q}_S$, signaling the onset of decoherence and the thermalization of the QOS, which offers a proxy for the thermalization process. 

Formally, the QME can be defined as follows~\cite{PhysRevLett.133.010401}: Consider a physical quantity $\Delta S(\rho)$ that monotonically measures the distance of a system from equilibrium. For two initial states $\rho_1(0)$ and $\rho_2(0)$ of a quantum system, the QME occurs if (i) $\Delta S(\rho_1(0)) < \Delta S(\rho_2(0))$ and (ii) $\Delta S(\rho_1(\tau)) > \Delta S(\rho_2(\tau))$, $\forall \tau >t_M$, where $t_M$ is the Mpemba time. In that case, the decay rate of these off-diagonal elements is crucial for understanding the timescale of thermalization and the emergence of QME.

We now turn our attention to the reduced density matrix \( \hat{\rho}_S \) of QOS, and analyze the evolution of the off-diagonal elements between two charge eigenstates. These off-diagonal elements are
\begin{widetext}
\begin{eqnarray}
\bra{S^{q_1}_{i_1}}\hat{\rho}_S(t)\ket{S^{q_2}_{i_2}} 
&& =  \bra{\Psi(t)} \left( \ket{S^{q_2}_{i_2}} \bra{S^{q_1}_{i_1}} \otimes \hat{I}_B \right) \ket{\Psi(t)} \nonumber\\
&& = \sum_{m_1, m_2, n_1, n_2} \braket{\Psi(0)|m_2, n_2} \bra{m_2, n_2}\left( \ket{S^{q_2}_{i_2}}\bra{S^{q_1}_{i_1}} \otimes \hat{I}_B \right) \ket{m_1, E_{n_1}} \nonumber\\
&& ~~~~\times \braket{m_1, E_{n_1}|\Psi(0)} \exp \left( -i \left(E^{m_1}_{n_1} - E^{m_2}_{n_2}\right) t \right) \nonumber\\
&& = \sum_{m, n_1, n_2} \bra{q_2 + m, E_{n_2}} \left( \ket{S^{q_2}_{i_2}} \bra{S^{q_1}_{i_1}} \otimes \hat{I}_B \right) \ket{q_1 + m, E_{n_1}} \nonumber\\
&& ~~~~\times \braket{q_1 + m, E_{n_1} | \Psi(0)} \braket{\Psi(0) | q_2 + m, E_{n_2}} \exp \left( -i \left(E^{m_2}_{n_2} - E^{m_1}_{n_1}\right) t \right).
\label{eq:off-diagonal}
\end{eqnarray}
\end{widetext}

The above expression involves a sum over all possible states $m_1, m_2$ in the Hilbert space, and the coefficients $\bra{m_2, n_2} \left( \ket{S^{q_2}_{i_2}} \bra{S^{q_1}_{i_1}} \otimes \hat{I}_B \right) \ket{m_1, E_{n_1}}$ are constrained by the conserved charge. Specifically, conservation of charge implies that
\begin{eqnarray}
\bra{m_2, n_2} \left( \ket{S^{q_2}_{i_2}}\bra{S^{q_1}_{i_1}} \otimes \hat{I}_B \right) \ket{m_1, E_{n_1}} \propto \delta_{m_1 - q_1,m_2 - q_2}. \nonumber\\
\end{eqnarray}
which induces the vairable substitution in Eq.~\ref{eq:off-diagonal} as $m_1 - q_1=m_2 - q_2=m$.

This delta function constraint ensures that the system evolution only couples states within the same charge sector, reducing the summation to just a single charge subspace, which explains why only the charge index of the subspace remains relevant in the final expression.

Now, as a common and physically relevant assumption, we consider the case where the QOS and QTB are initially unentangled. In this scenario, the total initial state of the system can be written as
\begin{eqnarray}
\Ket{\Psi(0)} = \left( \sum_{q, i} c_i^q \Ket{S_i^q} \right) \otimes \Ket{\psi_B},
\end{eqnarray}
where $\ket{\psi_B}$ is the initial state of the thermal bath. This leads to the simplification of the expression for the off-diagonal element as follows:
\begin{widetext}
\begin{eqnarray}
\bra{S^{q_1}_{i_1}}\hat{\rho}_S(t)\ket{S^{q_2}_{i_2}} &&= \sum_{q'_{1,2}c'_{1,2}}\sum_{m,n_1,n_2} c^{q'_1}_{c'_1}c^{q'_2*}_{c'_2}\bra{q_2+m, E_{n_2}}\left( \ket{S^{q_2}_{i_2}}\bra{S^{q_1}_{i_1}} \otimes \hat{I}_B \right)\ket{q_1+m,E_{n_1}} \nonumber\\
&& ~~~~\times\bra{q_1+m,E_{n_1}}\left(\ket{S^{q'_1}_{i'_1}}\bra{S^{q'_2}_{i'_2}}\otimes\ket{\psi_B} \bra{\psi_B}\right)\ket{q_2+m, E_{n_2}}\nonumber\exp(-i(E^{m+q_2}_{n_2}-E^{m+q_1}_{n_1})t) \nonumber\\
&&= \sum_{q'_{1,2}c'_{1,2}}\sum_{m,n_1,n_2} c^{q'_1}_{c'_1}c^{q'_2*}_{c'_2}\bra{q_2+m, E_{n_2}}\left( \ket{S^{q_2}_{i_2}}\bra{S^{q_1}_{i_1}} \otimes \hat{I}_B \right)\ket{q_1+m,E_{n_1}} \nonumber\\
&&~~~~\times\bra{q_1+m,E_{n_1}}\left(\ket{S^{q'_1}_{i'_1}}\bra{S^{q'_2}_{i'_2}}\otimes\ket{\psi^{q_1+m-q'_1}_B} \bra{\psi^{q_2+m-q'_2}_B}\right)\ket{q_2+m, E_{n_2}} \exp(-i(E^{m+q_2}_{n_2}-E^{m+q_1}_{n_1})t)\nonumber\\
\label{eq:off-diagonal-simplify}
\end{eqnarray}
\end{widetext}
where the identity operator of QTB has been ignored due to simplicity. In this expression, we have used the fact that the initial state of the thermal bath can be expanded in terms of the charge sectors of the QTB. Specifically, the thermal bath's state $|\psi_B\rangle$ can be projected onto the $q$-th subspace using the projection operator $\hat{\Pi}^q$, such that the state in the $q$-th subspace is $|\psi_B^{q}\rangle = \hat{\Pi}^q |\psi_B \rangle$.

We observe that \( |\psi^{q_1 + m - q'_1}_B \rangle \langle \psi^{q_2 + m - q'_2}_B | \) represents distinct block matrices in the initial density matrix of the QTB. This suggests that, when analyzing the evolution of the off-diagonal elements of the QOS density matrix, the contribution of each block can be considered independently, with each block's contribution being additive in nature. In the following, we explore the effect of each block on the evolution of the QOS density matrix.

We begin with a simple case where the QTB is assumed to be in a post-thermalized state. In the absence of other symmetries, the thermalization of this system constitutes a chaotic dynamical evolution, which can be viewed as a scrambling process. It is appropriate to express the initial state of the QTB as~\cite{PhysRevLett.126.190601,PhysRevLett.128.060601,cotler2017chaos,d2016quantum,rigol2008thermalization}

\begin{eqnarray}
    \left\langle|\psi_B\rangle\langle\psi_B|\right\rangle &&= \left\langle\hat{U}_B(t) |\psi_B\rangle\langle\psi_B| \hat{U}_B^{\dagger}(t)\right\rangle_t \nonumber\\
    &&= \sum_q \frac{p_q}{D_q} \hat{I}^q_B
\end{eqnarray}
where \( \hat{U}_B = \exp(-i \hat{H}_B t) \) represents the unitary evolution operator of the QTB, with \( \hat{H}_B \) being the Hamiltonian of the thermal bath. Here, \( \hat{I}^q_B \) is the identity operator acting on the \( q \)-th charge sector of the QTB, and \( p_q \) denotes the probability distribution over these charge sectors, $D_q = C(L_B,q)$ is the dimension of $\hat{I}^q_B$.

The averaging over time in the above expression is introduced for analytical simplicity and to facilitate subsequent theoretical derivations. This trick is also used by other works~\cite{paviglianiti2024enhanced}. However, it is important to note that this step is not strictly necessary for the emergence of the phenomena discussed later in this work. Numerical simulations and experimental observations confirm that the key results and physical insights remain valid even in the absence of such averaging. This robustness underscores the generality of the derived conclusions, which are not contingent on the specific assumption of time-averaging in the initial state preparation.

Using this, we can simplify the expression for the reduced density matrix of the QOS. Specifically, we substitute the post-thermalized QTB state into the expression Eq.~\ref{eq:off-diagonal-simplify}, yielding
\begin{widetext}
\begin{eqnarray}
\langle S^{q_1}_{i_1}|\hat{\rho}_S(t)|S^{q_2}_{i_2}\rangle && = \sum_{q',i'_{1,2}}\sum_{m,n_1,n_2} \frac{p_{m-q'}}{D_{m-q'}}c^{q_1+q'}_{i'_1}c^{q_2+q'*}_{i'_2}\langle q_2+m, E_{n_2}|\left( |S^{q_2}_{i_2}\rangle\langle S^{q_1}_{i_1}| \otimes \hat{I}^{m}_B \right)|q_1+m,E_{n_1}\rangle\nonumber\\
&&~~~~\times\langle q_1+m,E_{n_1}|\left(|S^{q_1+q'}_{i'_1}\rangle\langle S^{q_2+q'}_{i'_2}|\otimes\hat{I}^{m-q'}_B\right)|q_2+m, E_{n_2}\rangle \exp(-i(E^{m+q_2}_{n_2}-E^{m+q_1}_{n_1})t)
\label{eq:off-diagonal-2}
\end{eqnarray}
\end{widetext}
This expression implies that for a given initial state of the QTB, the coherence between different subspaces of the QTB can indeed influence the off-diagonal elements of the QOS density matrix. However, when averaging over all possible initial states of the QTB, the contributions arising from this entanglement asymmetry effectively vanish. 

Furthermore, it can be demonstrated that the term corresponding to 
$q'=0$ contributes dominantly to the summation, see Appendix.~\ref{appendix:neglect}. This conclusion is further supported by numerical simulations. As a result, the evolution of the off-diagonal elements of the QOS density matrix can be expressed as
\begin{widetext}
\begin{eqnarray}
\langle S^{q_1}_{i_1}|\hat{\rho}_S(t)|S^{q_2}_{i_2}\rangle && = \sum_{i'_{1,2}}\sum_{m,n_1,n_2} \frac{p_m}{D_m}c^{q_1}_{i'_1}c^{q_2*}_{i'_2}\langle q_2+m, E_{n_2}|\left( |S^{q_2}_{i_2}\rangle\langle S^{q_1}_{i_1}| \otimes \hat{I}^{m}_B \right)|q_1+m,E_{n_1}\rangle\nonumber\\
&&~~~~\times\langle q_1+m,E_{n_1}|\left(|S^{q_1}_{i'_1}\rangle\langle S^{q_2}_{i'_2}|\otimes\hat{I}^{m}_B\right)|q_2+m, E_{n_2}\rangle\exp\left(-i(E^{m+q_2}_{n_2}-E^{m+q_1}_{n_1})t\right)
\label{eq:off-diagonal-3}
\end{eqnarray}
\end{widetext}

At this stage, the evolution of the off-diagonal elements of the QOS can be interpreted as a linear superposition of decaying curves, each contributed by distinct subspaces of the QTB indexed by $m$, just as illustrated in Fig.~\ref{fig:demo}. These curves are effectively decoupled from one another. Moreover, when we consider how the initial state of QOS influence the evolution of $\langle S^{q_1}_{i_1}|\hat{\rho}_S(t)|S^{q_2}_{i_2}\rangle$, we could find it to be primarily governed by the initial coherence between the $q_1$-th and $q_2$-th subspaces within the QOS, which is reflected by $c^{q_1}_{i'_1}c^{q_2*}_{i'_2}$, appears to be coefficients to decide the curve amplitude. This implies that the relaxation velocity is predominantly influenced by the initial state of the QTB, or more detailed, the probability $\{p_m\}$ in different subspace, who could result in different relaxation time scales. The validity of this conclusion will be further substantiated by subsequent numerical results.

It is noteworthy that the decay rates of the aforementioned terms vary with 
$m$, which can be attributed to the effective temperature differences among the subspaces of the QTB. Back to \eqref{eq:off-diagonal-3}, the decay process contributed by the \( m \)-th subspace of the QTB is
\begin{widetext}
\begin{eqnarray}
&& \sum_{i'_{1,2}}\sum_{n_1,n_2} \frac{p_m}{D_m}c^{q_1}_{i'_1}c^{q_2*}_{i'_2}\langle q_2+m, E_{n_2}|\left( |S^{q_2}_{i_2}\rangle\langle S^{q_1}_{i_1}| \otimes \hat{I}^{m}_B \right)|q_1+m,E_{n_1}\rangle \nonumber\\
&& \times\langle q_1+m,E_{n_1}|\left(|S^{q_1}_{i'_1}\rangle\langle S^{q_2}_{i'_2}|\otimes\hat{I}^{m}_B\right)|q_2+m, E_{n_2}\rangle \exp\left(-i(E^{m+q_2}_{n_2}-E^{m+q_1}_{n_1})t\right).
\label{eq:off-diagonal-4}
\end{eqnarray}

As the energy levels become dense, this expression can be approximated as a continuous spectrum
\begin{eqnarray}
&& \sum_{n_1,n_2} \langle q_2+m, E_{n_2}|\left( |S^{q_2}_{i_2}\rangle\langle S^{q_1}_{i_1}| \otimes \hat{I}^{m}_B \right)|q_1+m,E_{n_1}\rangle \nonumber\\
&& \times\langle q_1+m,E_{n_1}|\left(|S^{q_1}_{i'_1}\rangle\langle S^{q_2}_{i'_2}|\otimes\hat{I}^{m}_B\right)|q_2+m, E_{n_2}\rangle \exp\left(-i(E^{m+q_2}_{n_2}-E^{m+q_1}_{n_1})t\right) \nonumber\\
 && = \int_{\omega=-\infty}^{\infty} N(\omega)M(\omega) \exp(-i\omega t) \, d\omega,
\end{eqnarray}
where \( \omega = E^{m+q_2}_{n_2} - E^{m+q_1}_{n_1} \), and \( M(\omega) \) represents the expectation value
\begin{eqnarray}
M(\omega) &&= \left\langle \langle q_2+m, E_{n_2} | 
\left( |S^{q_2}_{i_2} \rangle \langle S^{q_1}_{i_1}| \otimes \hat{I}^{m}_B \right) 
| q_1+m, E_{n_1} \rangle \right. \nonumber\\
&&\quad \times \left. \langle q_1+m, E_{n_1} | 
\left( |S^{q_1}_{i'_1} \rangle \langle S^{q_2}_{i'_2} | \otimes \hat{I}^{m}_B \right) 
| q_2+m, E_{n_2} \rangle 
\right\rangle \Big|_{E_{n_2}-E_{n_1}=\omega},
\end{eqnarray}
\end{widetext}
in which the Eigenstate Thermalization Hypothesis (ETH) ensurence that $M(\omega)$ is continuous function of $\omega$~\cite{d2016quantum}. Here, \( N(\omega) \) denotes the density of states pairs with energy gap \( \omega \). It is observed that for different values of \( m \), \( N(\omega) \) exhibits varying decay rates as \( \omega \) deviates from the central value, which would also be illustrated by Fig.~\ref{fig:spectrum} in an exact model introducted later. This results in different spectral widths for the considered term, depending on \( m \). The spectral width reflects the characteristic time scale, thereby explaining the variation in decay rates across subspaces indexed by \( m \).

To further elucidate the origin of this difference, we consider the variance of the energy gap distribution, \( Var(\omega) \), which characterizes the spectral width of \( N(\omega) \):
\begin{equation}
Var(\omega) = Var\left(E^{m+q_1}_{n_1}\right) + Var\left(E^{m+q_2}_{n_2}\right),
\end{equation}
where we have assumed that the energy levels within the \( m+q_1 \) and \( m+q_2 \) subspaces are statistically independent, as they belong to distinct symmetry sectors. 

This expression reveals that a broader eigenenergy distribution within the relevant subspaces leads to a larger variance \( Var(\omega) \), and consequently to a broader spectral width for the term Eq.\ref{eq:off-diagonal-4}. Since the spectral width is inversely related to the characteristic relaxation timescale, a greater variance directly corresponds to faster relaxation dynamics.

Therefore, to enhance the relaxation process, one should identify pairs of subspaces \( (m+q_1, m+q_2) \) that exhibit more dispersed eigenenergy distributions, i.e., larger values of \( Var\left(E^{m+q_1}_{n_1}\right) + Var\left(E^{m+q_2}_{n_2}\right) \). Once such favorable subspaces are determined, increasing the occupation probability within these sectors would effectively accelerate the relaxation of the quantum open system (QOS), as will be demonstrated in subsequent sections. The connection between the QME and the distribution of occupation probabilities across different subspaces has also been noticed in earlier work~\cite{PhysRevLett.133.140405}, lending further support to the conclusions drawn in this study.

We could provide a thermodynamic interpretation linking spectral properties of quantum chaotic systems to their thermalization dynamics.

A key observation is that a more dispersed energy spectrum corresponds to an enhanced local density of states within any given energy window. Formally, the microcanonical entropy is related to the density of states $\Omega(E)$ through
\begin{eqnarray}
    S(E) = k_B \ln \Omega(E),
\end{eqnarray}
where $k_B$ denotes the Boltzmann constant. A higher density of states implies a steeper entropy growth with energy, leading to a faster expansion of the accessible Hilbert space volume during dynamical evolution.

From a dynamical standpoint, when the initial state is expanded over a large number of closely spaced energy eigenstates, the time evolution involves rapidly varying relative phases, resulting in quick dephasing and equilibration of local observables. This mechanism accelerates the apparent thermalization process, in agreement with the ETH.

Moreover, drawing an analogy to classical thermodynamics, one can regard the effective dynamical landscape of the system: a higher density of states smoothens the "free energy landscape," facilitating efficient exploration of configuration space, while a compressed spectrum corresponds to a rugged or bottlenecked landscape, impeding relaxation.

\begin{figure*}[]
    \centering
    \includegraphics[]{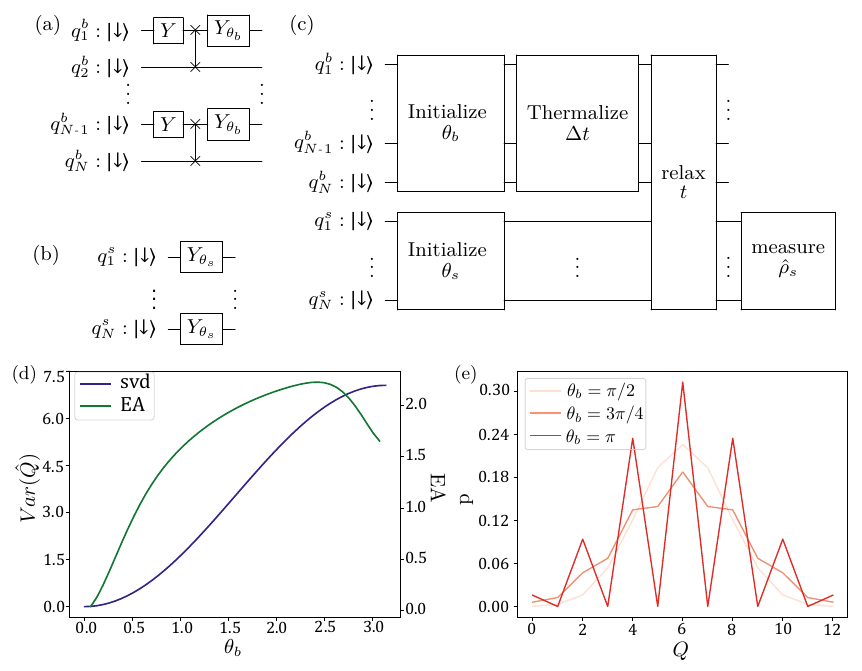}
    \captionsetup{justification=raggedright,singlelinecheck=false}
    \caption{(a) Quantum circuit for the initialization of QTB. The qubits, denoted as \( q^b_{i} \), are partitioned into pairs, with each pair prepared in an identical state determined by the parameter \(\theta_b\) as discussed in the main text. (b) Quantum circuit for the initialization of QOS. The qubits, labeled as \( q^s_i \), are rotated by the same angle \(\theta_s\), ensuring no entanglement is introduced between them. (c) Schematic of the experimental circuit. After the QTB evolves for a random duration $\Delta t$, the coupling between the QOS and QTB is activated. The reduced density matrix of the QOS, \(\rho_s(t)\), is subsequently measured at time \( t \) following the onset of relaxation dynamics. (d) Variance of conserved charge \(Var(\hat{Q}) = \langle \hat{Q}^2 \rangle - \langle \hat{Q} \rangle^2 \) and the entanglement asymmetry $\Delta S$ of the QTB under different $\theta_b$. (e) Occupation probability in different eigenspace of conserved charge $\hat{Q}$ of the QTB.}
    \label{fig:quantum_circuits}
\end{figure*}

\section{Numerical Verification}
To construct a non-integrable model with only a locally conserved charge as the symmetry, we consider a one-dimensional XY model with open boundary conditions, composed of spin-1/2 particles. The Hamiltonian is given by
\begin{eqnarray}
\hat{H} = &&\sum_{i<j} \frac{J}{|i-j|} \left( \sigma^x_i \sigma^x_j + \sigma^y_i \sigma^y_j \right) + \nonumber\\
&&h \sum_{i=1}^N \left( i - \frac{L}{2} \right) \sigma^z_i,
\end{eqnarray}
where \( L \) denotes the length of the spin chain. This model has been established in some recent experimental work~\cite{joshi2024observing,PhysRevA.99.052342,RevModPhys.95.035002}. A linear longitudinal field is introduced to break the \( S_2 \) symmetry. This Hamiltonian is verified to be non-integrable, with the only conserved charge being \( \hat{Q} = \sum_i \hat{\sigma}^z_i \)~\cite{giraud2022probing}, which can be decomposed into the sum of charges from two subsystems. Here, we set \( J = 1.0 \), \( h = 0.2 \), and the total system size \( L = 15 \), with \( L_b = 12 \) qubits constituting the quantum thermal bath (QTB) and \( L_s = 3 \) qubits forming the quantum open system (QOS), located at positions \( [7, 9] \).

\begin{figure*}[t]
    \centering
    \includegraphics[width=0.92\linewidth]{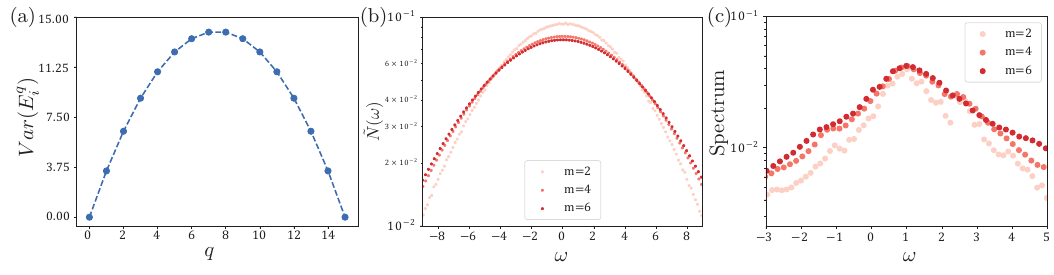}
    \captionsetup{justification=raggedright,singlelinecheck=false}
    \caption{(a) variance of energy distribution in the $q$-th subspace. (b) and (c) Analysis of the matrix element \( |S^{q_2}_{i_2}\rangle\langle S^{q_1}_{i_1}| = |S^{q_2}_{i'_2}\rangle\langle S^{q_1}_{i'_1}| = \ket{\downarrow\downarrow\uparrow}\bra{\downarrow\uparrow\uparrow} \). each point in the figure reflects a sum over the neiborhood area of frequency \(\delta\omega\) (b) Energy level density $\tilde{N}(\omega)=N(\omega)/dim_m$ for different \( m \), where $dim_m$ is the total number of levels. (c) Numerical scatter plots of \( N(\omega)M(\omega) \).}
    \label{fig:spectrum}
\end{figure*}

As discussed theoretically, each symmetry subspace of the QTB contributes a distinct relaxation rate. Analysis of the eigensystem structure reveals that, when considering the influence of the \( m \)-th subspace of the QTB, the closer \( \frac{q_1 + q_2}{2} + m \) is to \( \frac{L}{2} \), the faster the decoherence between the \( q_1 \)-th and \( q_2 \)-th subspaces of the QOS. Numerical evidence has shown in Fig.~\ref{fig:spectrum}, in which we could find when $m$ steps over ${2,4,6}$, the width of $N(\omega)$ and spectrum would be larger and larger, which implies a faster relaxation dynamics.

\begin{figure}
    \centering
    \includegraphics[width=0.96\linewidth]{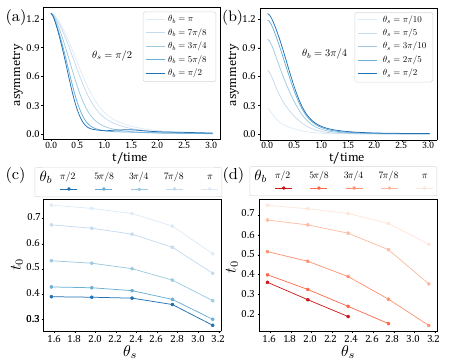}
    \captionsetup{justification=raggedright,singlelinecheck=false}
    \caption{(a) Numerical results demonstrating the relaxation dynamics of QOS initialized to the same state ($\theta_s = \pi/2$) under varying initial states of QTB. The initialization parameters of the QTB are denoted by $\theta_b$. (b) Relaxation behavior of the QOS with different initial states ($\theta_s$) while the QTB is fixed in the same initial state ($\theta_b = 3\pi/4$). (c) and (d) the decaying curve of asymmetry is fitted by function $y=A\exp(-t^2/t_0^2)$, and the parameter $t_0$ under different initial state is shown. (c) corresponds to the case of averaging over multiple samplings of $\Delta t$, and (d) corresponds to the case of $\Delta t=100$ without averaging. In (d) the curve of big $\theta_s$ and small $\theta_b$ showed significant deviations from the fitting model and were therefore excluded from the fitting analysis presented in the figure.}
    \label{fig:NumericalResult}
\end{figure}

\begin{figure*}
    \centering
    \includegraphics[width=0.66\linewidth]{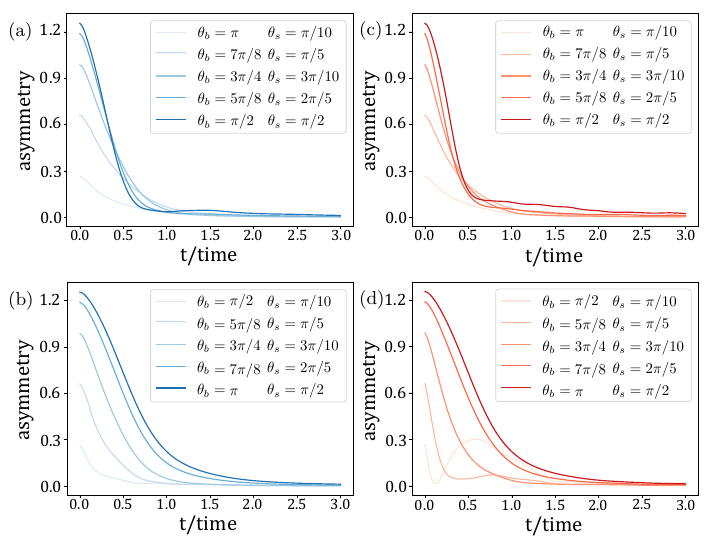}
    \captionsetup{justification=raggedright,singlelinecheck=false}
    \caption{(a) and (b), Investigation of QME by independently initializing the QOS and QTB to distinct states. (a) A case where QME is observed, characterized by the presence of a crossover point in the asymmetric relaxation curves. (b) A contrasting scenario where QME is absent, as evidenced by the lack of a clear crossover. (c) and (d), Relaxation dynamics with a thermalization time $\Delta t = 100$, presented without averaging over multiple samplings of $\Delta t$.}
    \label{fig:QMEhappenornot}
\end{figure*}    

To validate this effect, we design an initial state for the QTB as follows: the spins in the QTB are divided into pairs, with each pair initialized in the state
\begin{eqnarray}
\ket{\varphi_{B0}} = \frac{\sin(\theta_b/2)}{\sqrt{2}} \left( \ket{\uparrow\uparrow} + \ket{\downarrow\downarrow} \right) + \nonumber\\
\frac{\cos(\theta_b/2)}{\sqrt{2}} \left( \ket{\uparrow\downarrow} + \ket{\downarrow\uparrow} \right).
\end{eqnarray}
This state ensures that the expectation value of the symmetry charge remains at half-filling, while the variance of the symmetry charge can be tuned by adjusting \( \theta_b \). A higher variance corresponds to a greater occupation probability in non-half-filled subspaces, which is expected to slow down the relaxation of the QOS. Additionally, this design is experimentally feasible, as demonstrated by the quantum circuit shown in Fig.~\ref{fig:quantum_circuits}. One interesting thing here is that the initial entanglement asymmetry is not a monotonic functions of \( \theta_b \), which means in this model initial asymmetric of QTB is not a main factor that influence the relaxation velocity.

For the initialization of the QOS, a tilted Néel state is employed
\begin{eqnarray}
|\varphi_{S0}\rangle = \exp\left( -i \frac{\theta_s}{2} \sum_{i \in \text{QOS}} \hat{\sigma}^y_i \right) \ket{\downarrow\dots\downarrow}.
\end{eqnarray}

The experimental circuit, illustrated in Fig.~\ref{fig:quantum_circuits}, includes a random thermalization time \( \Delta t \). The reduced density matrix of the QOS, \( \tilde{\hat{\rho}}_s \), is obtained by averaging over multiple thermalization times \( \Delta t \)
\begin{eqnarray}
\tilde{\hat{\rho}}_s = \frac{1}{N_s} \sum_{\Delta t_i}^{N_s} \hat{\rho}_s,
\end{eqnarray}
where \( N_s = 100 \) is the number of samples. The relaxation dynamics of the QOS are evaluated using a recently induced measure, entanglement asymmetry (EA)~\cite{ares2023entanglement,10.21468/SciPostPhys.15.3.089,ferro2024non,capizzi2023entanglement}, defined as the quantum relative entropy between the reduced density matrix \( \rho_A(t) \) and its symmetrized version \( \rho_{A,Q} \). Here, \( \rho_{A,Q} \) is obtained by projecting \( \rho_A(t) \) onto the eigenspaces of the symmetry operator \( Q_A \)
\begin{eqnarray}
\rho_{A,Q} = \sum_q \Pi_q \rho_A(t) \Pi_q,
\end{eqnarray}
where \( \Pi_q \) is the projector onto the eigenspace of \( Q_A \) with eigenvalue \( q \). The entanglement asymmetry is expressed as
\begin{eqnarray}
\Delta S_{A}(t) = \mathrm{tr}\left[ \rho_A(t) \left( \log \rho_A(t) - \log \rho_{A,Q}(t) \right) \right].
\end{eqnarray}

The results, shown in Fig.~\ref{fig:NumericalResult}, demonstrate that the relaxation rate of the QOS varies significantly with different initial states of the QTB. As predicted, a lower charge variance in the QTB leads to faster relaxation. Conversely, when the QTB is initialized in the same state, variations in the initial state of the QOS have little impact on the relaxation rate, consistent with our theoretical analysis.

With this finding, we can systematically investigate the conditions under which the QME manifests or fails to occur. This is achieved by examining various combinations of initial QOS states coupled with different QTB configurations. As illustrated in Fig.~\ref{fig:QMEhappenornot}, our analysis reveals that the QME phenomenon emerges when a QOS initialized far from its equilibrium state interacts with a QTB capable of inducing rapid thermal relaxation. Conversely, the QME effect is suppressed when these specific conditions are not met. This dichotomous behavior underscores the critical dependence of the QME on both the initial non-equilibrium character of the QOS and the relaxation dynamics governed by the QTB properties.

Furthermore, even without averaging over different thermalization times \( \Delta t \), our method remains effective in controlling the occurrence of QME. The primary difference lies in the presence of fluctuations in the relaxation dynamics, as previously discussed in the theoretical section.

\section{Conclusion}
In this work, we have investigated the influence of the initial state of the QTB on the relaxation dynamics of a QOS within a fully Schrödinger-evolved quantum framework, without invoking quasiparticle or dynamical mode approximations. Our results reveal that the occupation probability distribution across different symmetry subspaces of the QTB critically influences the relaxation behavior of the QOS. Importantly, we establish that the relaxation rate could be governed by the spectral dispersion within each subspace: broader spectral distributions promote faster entropy production and more efficient thermalization dynamics. Furthermore, we demonstrate that by controlling the initial charge distribution across the symmetry subspaces of the QTB, it is possible to effectively modulate the relaxation velocity of the QOS. 

We proposed an experimentally feasible setup to validate our theoretical predictions, which successfully demonstrated that the relaxation dynamics of the QOS are significantly influenced by the initial state of the QTB. Our results indicate that a lower charge variance in the QTB leads to faster relaxation of the QOS, while variations in the initial state of the QOS itself have minimal impact on the relaxation rate. This aligns with our theoretical analysis, which suggests that the initial state of the QTB primarily governs the relaxation dynamics, whereas the initial state of the QOS mainly sets the starting point of its evolution.

Furthermore, our work highlights the potential for controlling the occurrence of the QME by engineering the initial state of the QTB and QOS. This opens up new avenues for manipulating non-equilibrium quantum systems, with potential applications in quantum computing and quantum thermodynamics. Importantly, we have also established a connection between the relaxation dynamics and the spectral properties of the system. By analyzing the energy level density and spectral dispersion within different symmetry subspaces, we linked the relaxation rates to the distribution of spectral widths in the symmetry subspaces. This provides a novel perspective for understanding quantum non-equilibrium dynamics, as it bridges the gap between microscopic quantum evolution and macroscopic thermodynamic behavior. This connection not only deepens our theoretical understanding but also contributes to the broader field of quantum thermodynamics, shedding new light on how quantum systems approach equilibrium.

However, there remain several areas for further exploration. For instance, the detailed mechanisms beyond the occupation probability distribution in different symmetry subspaces of the QTB warrant further investigation. Additionally, the role of other symmetries and their interplay with the conserved charge could provide further insights into the relaxation dynamics of quantum systems.

In conclusion, our study provides a comprehensive understanding of how the initial state of the QTB influences the relaxation of the QOS, which promises a approach for controlling quantum relaxation processes in non-equilibrium systems. This work not only advances our theoretical understanding but also paves the way for practical applications in quantum technologies. By connecting these dynamics to thermodynamic quantities, we have laid the groundwork for future explorations in quantum thermodynamics, leaving an intriguing contribution to the field.

\begin{acknowledgments}
    We thank Ziyong Ge for the helpful discussions and the support from the Synergetic Extreme Condition User Facility (SECUF) in Huairou District, Beijing. This work was supported by the National Natural Science Foundation of China (Grants Nos.~92265207, T2121001, 11934018, T2322030, 12122504, 12274142, 12475017), the Beijing Natural Science Foundation (Grant No.~Z200009), the Innovation Program for Quantum Science and Technology (Grant No.~2021ZD0301800), the Beijing Nova Program (No.~20220484121), the Scientific Instrument Developing Project of Chinese Academy of Sciences (Grant No.~YJKYYQ20200041), 
    the Natural Science Foundation of Guangdong Province (Grant No.~2024A1515010398), the Nippon Telegraph and Telephone Corporation (NTT) Research, the Japan Science and Technology Agency (JST)
        [via the Quantum Leap Flagship Program (Q-LEAP), and the Moonshot R\&D Grant Number JPMJMS2061], the Asian Office of Aerospace Research and Development (AOARD) (via Grant No. FA2386-20-1-4069), and the Office of Naval Research (ONR). 
\end{acknowledgments}

\appendix
\section{neglect of terms with different symmetry number}
\label{appendix:neglect}
As established in Eq.~\eqref{eq:off-diagonal-2} of the main text, the critical term governing symmetry-violating processes takes the form:
\begin{eqnarray}
\sum_{n_1,n_2,m} && \langle q_2+m, E_{n_2}|S^{q_2}_{i_2}\rangle\langle S^{q_1}_{i_1}|q_1+m,E_{n_1}\rangle \times \nonumber\\
&& \langle q_1+m,E_{n_1}|S^{q_1+q'}_{i'_1}\rangle\langle S^{q_2+q'}_{i'_2}|q_2+m, E_{n_2}\rangle \times \nonumber\\
&& \exp(-i(E^{m+q_2}_{n_2}-E^{m+q_1}_{n_1})t).
\end{eqnarray}
This term exhibits significant suppression when $q' \neq 0$, as will be rigorously demonstrated through Krylov space analysis. To elucidate this phenomenon, we first recast the expression in terms of time-ordered correlation functions
\begin{eqnarray}
\mathcal{C}(t) = \left\langle \hat{U}^\dagger(t) |S^{q_2}_{i_2}\rangle\langle S^{q_1}_{i_1}| \hat{U}(t) |S^{q_1+q'}_{i'_1}\rangle\langle S^{q_2+q'}_{i'_2}| \right\rangle_{\beta=0},
\end{eqnarray}
where $\langle\cdot\rangle_{\beta=0}$ denotes the infinite temperature ensemble average, and the bath identity operators have been suppressed for notational clarity. Introducing the operator pair
\begin{eqnarray}
    \hat{S}^{(1)} = |S^{q_2}_{i_2}\rangle\langle S^{q_1}_{i_1}|, \quad \hat{S}^{(2)} = |S^{q_1+q'}_{i'_1}\rangle\langle S^{q_2+q'}_{i'_2}|,
\end{eqnarray}
we recognize $\mathcal{C}(t)$ as a two-point time-ordered correlation function
\begin{eqnarray}
\mathcal{C}(t) = \left\langle \hat{U}^\dagger(t) \hat{S}^{(1)} \hat{U}(t) \hat{S}^{(2)} \right\rangle_{\beta=0}.
\end{eqnarray}
The dynamical properties of such correlation functions can be systematically analyzed through the lens of Krylov complexity~\cite{parker2019universal,barbon2019evolution,dymarsky2020quantum,rabinovici2021operator,caputa2022geometry}. For a generic Hamiltonian $\hat{H}$ and initial operator $\hat{O}$, the Krylov basis $\{\hat{S}_n\}_{n=0}^{D-1}$ is constructed via the Lanczos algorithm
\begin{eqnarray}
\hat{S}_0 &&= \hat{S}^{(1)}/\|\hat{S}^{(1)}\|, \nonumber\\
\hat{A}_1 &&= [\hat{H},\hat{S}_0],  b_1 = \|\hat{A}_1\|,  \hat{S}_1 = \hat{A}_1/b_1, \nonumber\\
\hat{A}_n &&= [\hat{H},\hat{S}_{n-1}] - b_{n-1}\hat{S}_{n-2}, b_n = \|\hat{A}_n\|,  \hat{S}_n = \hat{A}_n/b_n, \nonumber\\
\label{eq:phi_dynamics}
\end{eqnarray}
where $b_n$ are the Lanczos coefficients and $D$ is the Krylov space dimension. This orthogonalization procedure generates a tridiagonal representation of the Liouvillian superoperator $\mathcal{L} = [\hat{H},\cdot]$ under the inner product $(\hat{O}_1,\hat{O}_2) := \mathrm{Tr}[\hat{O}_1^\dagger\hat{O}_2]/\mathrm{Tr}[I]$ with induced norm $\|\hat{O}\| = (\hat{O},\hat{O})^{1/2}$.

The time-evolved operator admits a Krylov expansion
\begin{eqnarray}
\hat{S}^{(1)}(t) = e^{i\mathcal{L}t}\hat{S}^{(1)} = \sum_{n=0}^{D-1} i^n \phi_n(t) \hat{S}_n,
\end{eqnarray}
where the real-valued amplitudes $\phi_n(t)$ satisfy the recursive dynamics
\begin{eqnarray}
\partial_t \phi_n(t) = b_n \phi_{n-1}(t) - b_{n+1} \phi_{n+1}(t).
\end{eqnarray}
The correlation function consequently decomposes as
\begin{eqnarray}
\mathcal{C}(t) = \sum_{n=0}^{D-1} i^n c_n \phi_n(t), \quad c_n = (\hat{S}_n, \hat{S}^{(2)}).
\label{eq:correlation_decomposition}
\end{eqnarray}

The symmetry mismatch parameter $q'$ fundamentally controls the minimal Krylov depth $m$ required for non-vanishing overlap $c_n$. This arises because $\hat{S}^{(2)}$ resides in the symmetry sector shifted by $q'$ relative to $\hat{S}^{(1)}$, and each application of $\mathcal{L}$ (through the Lanczos algorithm) generates symmetry changes bounded by the Hamiltonian's hopping terms. Formally, there exists a minimal integer $m \propto |q'|$ such that $(\mathcal{L}^m \hat{S}^{(1)}, \hat{S}^{(2)}) \neq 0$. Since the Krylov basis elements satisfy $\hat{S}_n \in \mathrm{span}\{\mathcal{L}^k \hat{S}^{(1)}\}_{k=0}^n$, the decomposition \eqref{eq:correlation_decomposition} truncates to
\begin{eqnarray}
\mathcal{C}(t) = \sum_{n=m}^{D-1} i^n c_n \phi_n(t), \quad m \geq \lceil \alpha |q'| \rceil,
\end{eqnarray}
where $\alpha$ quantifies the symmetry change per Lanczos step.

Two key mechanisms enforce the suppression of $\mathcal{C}(t)$ for $q'\neq 0$:
\begin{enumerate}
    \item \textbf{Amplitude Delay:} Krylov wavefunctions $\phi_n(t)$ exhibit causal propagation through the basis. Higher $n$ thus imposes longer retardation timescales and diluted response with lower magnitude.~\cite{nandy2405quantum,parker2019universal}
    
    \item \textbf{Exponential Suppression:} In quantum chaotic systems characterized by linear Lanczos coefficient growth $b_n \sim \gamma n$~\cite{muck2022krylov,avdoshkin2020euclidean}, the overlap coefficients decay exponentially $c_n \sim e^{-\eta n}$ due to the orthogonality catastrophe in high-dimensional operator spaces. This can be shown from the iterative dynamics \eqref{eq:phi_dynamics}.
\end{enumerate}

\begin{figure}
    \centering
    \includegraphics[width=0.66\linewidth]{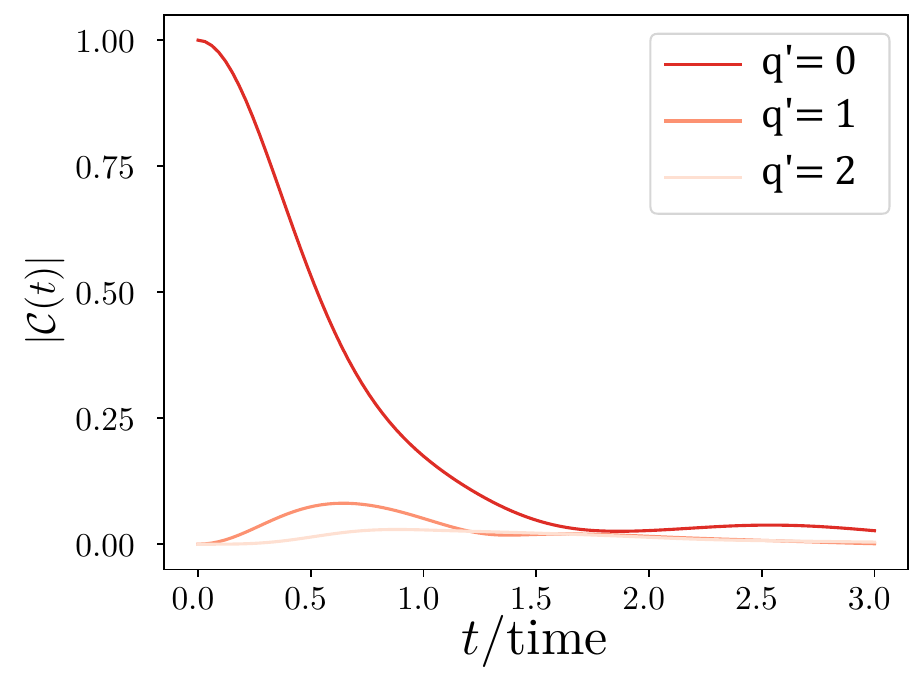}
    \captionsetup{justification=raggedright,singlelinecheck=false}
    \caption{Numerical results of correlation function $\mathcal{C}(t)$ for different $q'$.}
    \label{fig:correlation}
\end{figure}

Back to the numerical model in main text, we can also verify this by calculating the correlation function $\mathcal{C}(t)$ within different $q'$, results are shown in Fig.~\ref{fig:correlation}, in which
\begin{eqnarray}
    \hat{S}_1 &&= ~~\ket{\downarrow\downarrow\downarrow}\bra{\uparrow\downarrow\downarrow}\nonumber\\
    \hat{S}_2 &&= 
    \begin{cases}
        \ket{\uparrow\downarrow\downarrow}\bra{\downarrow\downarrow\downarrow} & q' = 0 \\
        \ket{\uparrow\uparrow\downarrow}\bra{\uparrow\downarrow\downarrow} & q' = 1 \\
        \ket{\uparrow\uparrow\uparrow}\bra{\uparrow\uparrow\downarrow} & q' = 2 \\
    \end{cases}.
\end{eqnarray}
From the numerical results, highly suppression of $\mathcal{C}(t)$ can be observed for $q' \neq 0$.

In summary, the symmetry mismatch parameter $q'$ controls the effective distance between operators $\hat{S}^{(1)}$ and $\hat{S}^{(2)}$ in Krylov space, with increasing $|q'|$ requiring progressively deeper exploration of the operator growth hierarchy. This geometric separation induces a dual suppression mechanism: correlations are both delayed in time and exponentially attenuated in amplitude due to the cumulative effects of Krylov complexity growth and quantum chaos. Consequently, terms with $q' \neq 0$ are subdominant at all relevant timescales, leaving the symmetry-preserving $q'=0$ component as the principal contributor to the dynamical evolution. This fundamental distinction justifies the perturbative treatment of $q' \neq 0$ terms in the main text's analysis.

\end{document}